\documentclass[prb,superscriptaddress, preprintnumbers ,twocolumn,showpacs,amsmath,amssymb]{revtex4}
\pdfoutput=1
\usepackage{amsmath}
\usepackage{amssymb}
\usepackage{amsthm}
\usepackage{amsfonts}
\usepackage{enumerate}
\usepackage{latexsym}
\usepackage{ifpdf}
\usepackage{graphicx}

\begin{document}

\title{Electronic and Magnetic Properties of the Interface between a
doped cuprate Y$_{0.6}$Pr$_{0.4}$Ba$_{2}$Cu$_{3}$O$_{7}$ and a
    colossal-magnetoresistance manganite
    La$_{2/3}$Ca$_{1/3}$MnO$_{3}$}
\author{Jian Liu}
\affiliation{Department of Physics, University of Arkansas, Fayetteville, Arkansas 72701,
USA}
\author{B. J. Kirby}
\affiliation{Center for Neutron Research, National Institute of Standards and Technology,
Gaithersburg, Maryland 20899, USA}
\author{M. Kareev}
\affiliation{Department of Physics, University of Arkansas, Fayetteville, Arkansas 72701,
USA}
\author{J. W. Freeland}
\affiliation{Advanced Photon Source, Argonne National Laboratory, Argonne, IL 60439, USA}
\author{H.-U. Habermeier}
\affiliation{Max Plank Institute for Solid State Research, D-70569 Stuttgart, Germany}
\author{G. Cristiani}
\affiliation{Max Plank Institute for Solid State Research, D-70569 Stuttgart, Germany}
\author{B. Keimer}
\affiliation{Max Plank Institute for Solid State Research, D-70569 Stuttgart, Germany}
\author{J. Chakhalian}
\affiliation{Department of Physics, University of Arkansas, Fayetteville, Arkansas 72701,
USA}

\begin{abstract}
The interfacial properties of Y$_{0.6}$Pr$_{0.4}$Ba$_{2}$Cu$_{3}$O$_{7}$/La$%
_{2/3}$Ca$_{1/3}$MnO$_{3}$ superlattices have been studied by resonant soft
x-rays and diffuse scattered neutrons. Linearly polarized X-ray absorption
at Cu L$_{3}$-edge reveals dramatic interfacial changes from the bulk
including charge transfer between LCMO and YPBCO and an orbital
reconstruction in the interface CuO$_{2}$ plane. The similarities to the
case of zero Pr-doping are due to the strongly hybridized covalent bond
between Cu and Mn. However, reduced charge transfer and a more bulk-like
interfacial orbital occupation are observed and related to the effect of
Pr-doping. Neutron reflectometry measurements reveal a drastic increase in
diffuse scattering with decreasing temperature, likely due to buckling
caused by the structural phase transition of the SrTiO$_{3}$ substrate. We
observe no evidence that this diffuse scattering is related to the
superconducting transition.
\end{abstract}

\date{\today}
\maketitle

Transition metal oxide (TMO) heterostructures have drawn enormous attention
because of the strongly correlated nature of their electronic behaviors. The
fine balance of the strong interactions between multiple degrees of freedom
can be changed and even manipulated at the interfaces to create new phases
that are not observed in the bulk materials. Heterostructures of high
temperature superconducting cuprates and colossal magnetoresistance
manganites are some of the most interesting oxide heterostructures due to
their incompatible order parameters (i.e. half-metallic ferromagnetism (FM) vs
superconductivity (SC)). In particular, the YBa$_{2}$Cu$_{3}$O$_{7}$ (YBCO) and La%
$_{2/3}$Ca$_{1/3}$MnO$_{3}$ (LCMO) heterostructures and superlattices are a
representative of these systems which have been intensively studied. High-Tc
SC was found to be suppressed by the presence of
the LCMO layer and dependent on the thickness of the LCMO layer.\cite%
{Sefrioui} Metallicity in both LCMO and YBCO layers was also shown to be
strongly suppressed with a large length scale.\cite{Holden} Furthermore, a
coupling of the superconducting layers was attributed to a long rang
proximity effect.\cite{Pena} On the other hand, X-ray Magnetic Circular
Dichroism (XMCD) did not reveal significant change of moments on manganese while
crossing the superconducting transition.\cite{Freeland1}

Recently, special interest has been focused onto the microscopic picture of
the YBCO/LCMO interface which also gives important insight into other
TMO interfaces in general. Specifically, the splitting of
the superlattice Bragg peak below the superconducting transition was
observed by off-specular neutron reflectometry (NR).\cite{Chakhalian1} More
surprisingly, XMCD at the Cu L-edge revealed
that the interfacial CuO$_{2}$ plane of YBCO has a net magnetic moment with
an unexpected antiferromagnetic coupling to Mn.\cite{Chakhalian1} This is
explained by the interfacial orbital reconstruction (thereafter OR) observed by X-ray Linear
Dichroism (XLD) at the Cu L-edge.\cite{Chakhalian2} Namely, while the holes
are constrained to Cu \textit{d}$_{x^{2}-y^{2}}$ orbital in bulk, a
significant number of holes occupy the \textit{d}$_{3z^{2}-r^{2}}$ orbital
at the interface. This is due to the strong covalent bond between Cu and Mn
\textit{d}$_{3z^{2}-r^{2}}$ orbitals causing the holes to move to the
antibonding state during charge transfer.

A very interesting remaining question is what role
superconductivity plays in YBCO/LCMO heterostructures. To address this
question, one needs to study the case when SC is suppressed.
One of the most interesting cases of suppressing High-Tc SC
is doping with Pr in the RBa$_{2}$Cu$_{3}$O$_{7}$ (123)
family (R= Y, Eu, Gd, etc.) because Pr is the only rare earth that maintains
the crystal structure throughout the entire doping range and suppresses
SC at the same time.\cite{Fehrenbacher,Mazin}
For example, Pr substitutes Y in the case of YBCO, and
SC is completely suppressed at as low as 50\% doping ratio.%
\cite{Peng} These properties are well suited for oxide heterostructures
since the oxygen stoichometry can be maintained. In addition, a giant
SC-induced modulation of the ferromagnetic magnetization from
one LCMO layer to the next one in a Pr-doped YBCO/LCMO superlattice has been
reported from NR lately.\cite{Hoppler1} The suppressed
SC was suggested to be an essential factor for the
surprisingly large double-period modulation of the vertical magnetic
profile, making Pr-doping especially interesting for the study of YBCO/LCMO
heterostructures. Therefore, in this article, we report our study on (Y,
Pr)BCO/LCMO superlattices which are similar to the previously studied
YBCO/LCMO\cite{Chakhalian1,Chakhalian2} but with 40\% Pr-doping to further
suppress SC.

Superlattices with a nominal structure of [Y$_{0.6}$Pr$_{0.4}$Ba$_{2}$Cu$_{3}%
$O$_{7}$ (100$%
\operatorname{\text{\AA}}%
)$/La$_{2/3}$Ca$_{1/3}$MnO$_{3}$ (100$%
\operatorname{\text{\AA}}%
)$]$_{5}$ were grown on 0.5mm-thick, atomically flat (001) SrTiO$_{3}$
(STO) single crystal substrates by pulsed laser deposition with a KrF
excimer Laser (248nm), the same as previously studied YBCO/LCMO
superlattices.\cite{Chakhalian1} The quality of the superlattices was checked by X-ray diffraction using Cu K$%
\alpha$ radiation (not shown). Only diffraction peaks due to
(Y,Pr)BCO and LCMO are observed, indicating a pure single phase. Moreover,
all (00\textit{l}) peaks are observed for both layers,
confirming the high-quality c-axis oriented epitaxial growth. The
corresponding (Y,Pr)BCO c-axis lattice parameter is around 11.66(3)$%
\operatorname{\text{\AA}}%
$ which is consistent with the reported value of the orthorhombic Y$_{1-x}%
$Pr$_{x}$Ba$_{2}$Cu$_{3}$O$_{7-\delta}$ structure.\cite{Peng}

Temperature dependent resistance was measured in a PPMS (Quantum Design)  with magnetic fields applied perpendicular to the film
surface. The result shown in Fig. 1 reveals the
well defined superconducting and magnetic behavior of the superlattice. At
zero field, the superconducting transition T$_{SC}$ (R=0) occurs
around 50 K which is much lower than the case without Pr-doping. However, it
is also slightly higher than the one corresponding to the nominal 40\%
Pr-doping ratio,\cite{Peng} indicating that the actual Pr concentration is
appreciably lower. As a field of 6 T is applied, T$_{SC}$ is suppressed to
around 30 K, and the transition is significantly broadened accompanied by a
positive magnetoresistance arised from vortex motion.\cite{Palstra} Compared
to the case of YBCO/LCMO with the field in the ab-plane,\cite{Chakhalian2}
the induced suppression and broadening by a field along c-axis are much more
pronounced and characteristic of superconducting cuprates due to the two
dimensional layered structure.\cite{Oda} At higher temperature, negative
magnetoresistance is observed resulting from the ferromagnetic state of
LCMO. Such an observation confirms the high-quality growth of both (Y,Pr)BCO
and LCMO layers as the their characteristic bulk behaviors are held.

\begin{figure}[t]\vspace{-0pt}
\includegraphics[width=7cm, height=4.75cm]{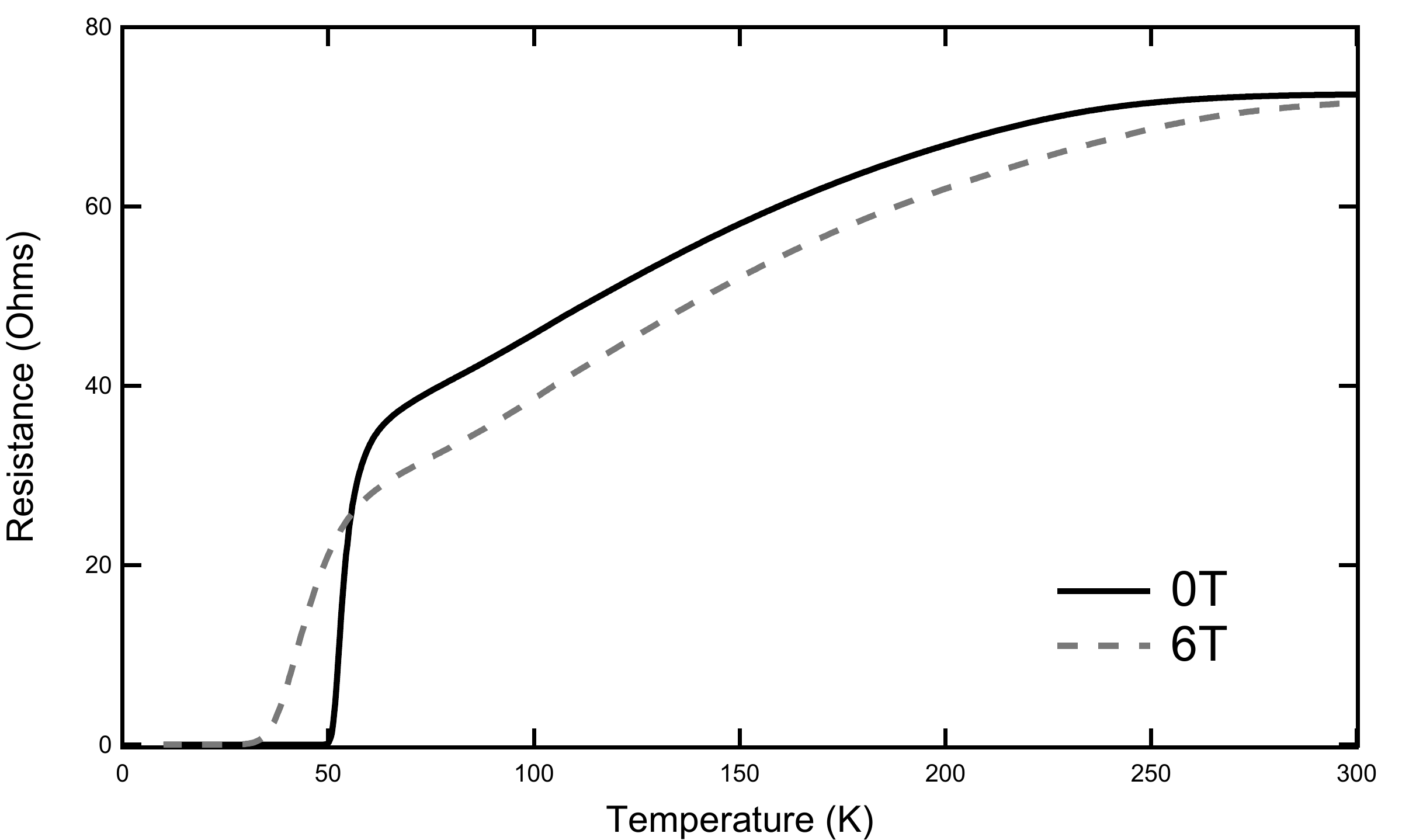}
\caption{\label{RvsT} Resistance vs temperature of [(Y,Pr)BCO (100$%
\operatorname{\text{\AA}}%
)$/ LCMO (100$%
\operatorname{\text{\AA}}%
)$]$_{5}$ superlattices without magnetic field (solid curve) and with a
magnetic field (dash curve) of 6T perpendicular to the film surface.}
\end{figure}

Linearly Polarized X-ray absorption experiments were performed at the 4-ID-C
beamline\cite{Freeland2} of the Advanced Photon Source at Argonne National
Laboratory. X-rays near Cu L$_{3}$-edge with polarizations along the c-axis
and in the ab-plane were used to obtain XLD. Absorption spectra were
recorded simultaneously in both fluorescence yield (FY) mode and total
electron yield (TEY) mode. These different modes provide different depth
sensitivities: FY mode is sensitive to the bulk of the superlattice, while
TEY mode probes the Cu state at the first interface covered by the top LCMO layer.\cite{Chakhalian2}

\begin{figure}[t]\vspace{-0pt}
\includegraphics[width=8cm, height=5.94cm]{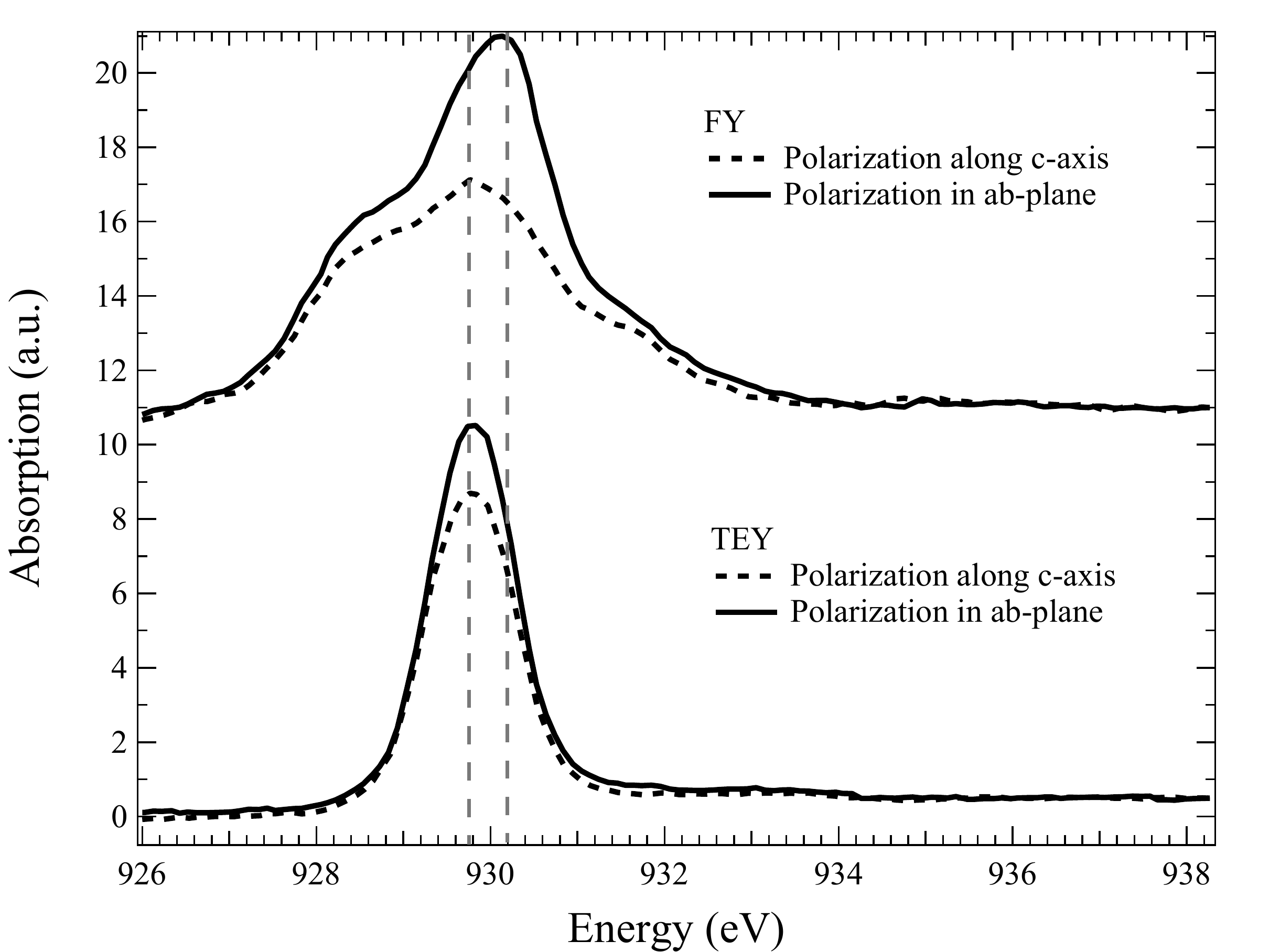}
\caption{Normalized X-ray absorption spectra at Cu L$_{3}$-edge taken in (a)
bulk-sensitive fluorescence mode and (B) interface-sensitive total electron
yield mode. Dash lines are guides to eyes.
   \label{XAS}}
\end{figure}

Figure 2 shows the normalized spectra of the Cu L$_{3}$-edge with in-plane
and out-of-plane polarizations. Figure 2 (a) and (b) corresponds to the
spectra in bulk-sensitive FY mode and interface-sensitive TEY mode,
respectively. The FY data of both polarizations show a main peak at 930.1eV,
a broad shoulder at lower energy and a relatively small shoulder at higher
energy. The main peak corresponds to the Cu 2\textit{p}$^{6}$3\textit{d}$%
^{9}\longrightarrow $2\textit{p}$^{5}$3\textit{d}$^{10}$ transition, while
the small shoulder is due to 2\textit{p}$^{6}$3\textit{d}$%
^{9}L\longrightarrow$2\textit{p}$^{5}$3\textit{d}$^{10}L$, where \textit{L}
denotes an oxygen ligand hole. Such a line shape of the Cu L$_{3}$-edge
absorption peak is the signature of the \textquotedblleft Zhang-Rice (ZR) singlet state\textquotedblright.\cite%
{Zhang} The shoulder on the left is due to the Pr M$_{5}$-edge which
partially overlaps with the Cu L$_{3}$-edge and, therefore, was not
resolved. This observation and the assignment above are consistent with
other reported X-ray absorption data of Pr-doped cuprates.\cite%
{Neukirch,Chen1,Chen2} In spite of these common features of the FY data of
both polarizations, it is clearly seen that the absorption of in-plane
polarization is much more intense than that of polarization along c-axis. In
particular, the difference reaches a maximum at the main peak but
significantly decreases and almost disappears when passing the Pr M$_{5}$%
-edge. Since, to our best knowledge, no XLD data at Pr M-edge is reported in
literature for Pr-doped cuprates to compare with, based on the observed
dichroism signal we conclude that there is no significant polarization
dependence at the Pr M-edge. Consequently, one can see that the Cu L$_{3}$%
-edge absorption of in-plane polarization is exceedingly stronger than that
of polarization along c-axis by taking the Pr shoulder as the reference.
Such an intense XLD of Cu L$_{3}$-edge implies that the holes on the Cu
d-shell predominantly occupy the planar \textit{d}$_{x^{2}-y^{2}}$ orbital,
which is also observed in all other high temperature superconductors.\cite%
{N¨¹cker,Chen3} In contrast to the bulk, the picture at the interface shown
by the interface-sensitive TEY data is dramatically different as can be seen
in Fig. 2(b). First of all, the small shoulder due to 2\textit{p}$^{6}$3%
\textit{d}$^{9}$\textit{L}$\longrightarrow$2\textit{p}$^{5}$3\textit{d}$^{10}
$\textit{L} transition in bulk is no longer present for either polarization,
implying that the ZR state is destroyed at the interface. Surprisingly, the
broad shoulder from Pr M$_{5}$-edge is also absent, by virtue of the electrostatic argument implying that it is more stable for Y ion to be located at the interfacial layer. More importantly, the
position of the main absorption peak is shifted by $\sim$0.3eV towards
lower energy, which is the sign of charge transfer across the interface
between two materials with different work functions. Compared to the Cu L$%
_{3}$-edge positions of other Cu valences, such a chemical shift corresponds
to a charge transfer less than 0.2\textit{e} per Cu ion.\cite{de Groot}
Furthermore, the polarization dependence of the absorption at the interface
is much weaker than in the bulk, which is the signature of the interfacial
OR. The strong enhancement of the absorption intensity
of polarization along c-axis illustrates that a large hole population
resides on the d$_{3z^{2}-r^{2}}$ orbital. As mentioned earlier, an analogue
of chemical shift and OR has also been observed at the
interface between optimally doped YBCO and LCMO.\cite{Chakhalian2} Although
SC is reduced by Pr-doping in the present case, the similar
interfacial modification from the bulk is due to the strong hybridization
nature of the covalent bonding between Cu and Mn.

\begin{figure}[t]\vspace{-0pt}
\includegraphics[width=8.5cm, height=3.77cm]{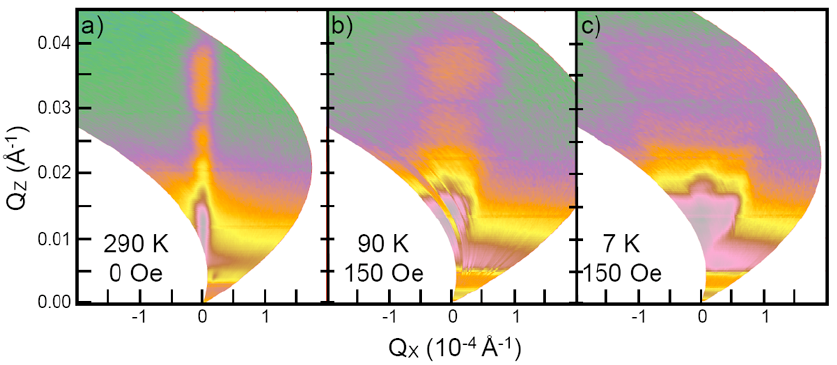}
\caption{(Color online) Off-specular neutron reflectivity taken at room temperature, 90K
and 7K with a 15 mT cooling field applied parallel to the layers and
perpendicular to the beam.
\label{Maps}}
\end{figure}

However, there are still differences between these two cases. First, the
amount of transferred charge is reduced in the present case given by the
smaller shift of the absorption peak. Furthermore, the interfacial orbital
occupation is more bulk-like evidenced by the relative intensity of the two
polarizations. The explanation of these features is most likely related to
the role played by Pr-doping. As is well known, at zero Pr-doping the ZR
singlet band lies across the Fermi level in between the upper Hubbard band
and the oxygen band. Although there is still controversy on how Pr-doping
affects SC in the 123 family, it is widely accepted that, as
the doping ratio increases, the so-called \textquotedblleft Fehrenbacher-Rice\textquotedblright\ band emerges
at the Fermi surface due to the strong hybridization between the O \textit{p}
state and the Pr \textit{f}$_{z(x^{2}-y^{2})}$ state.\cite{Fehrenbacher,Mazin,Merz} As a result, this emerging band
removes the holes from the ZR band and consequently raises the Fermi level.
Therefore, one would expect that the hole acquisition by LCMO at the
interface is reduced and, hence, the partial occupancy of the Cu \textit{d}$%
_{3z^{2}-r^{2}}$ orbital in the interfacial covalent bond decreases, as
reflected in the experimental results.

\begin{figure}[t]\vspace{-0pt}
\includegraphics[width=6cm, height=5.29cm]{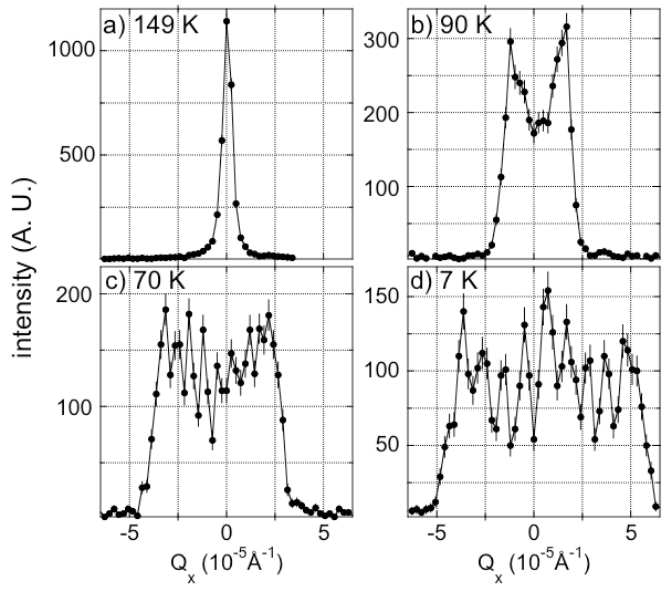}
\caption{Temperature-dependent unpolarized neutron transverse scans at
\textit{Q}$_{z}$ = 0.0139$%
\operatorname{\text{\AA}}%
^{-1}$ (right below the critical edge). Error bars correspond to $\pm$$\sigma$
. Lines are guides to eye.
\label{Rock}}
\end{figure}

\begin{figure}[t]\vspace{-0pt}
\includegraphics[width=5cm, height=4.5cm]{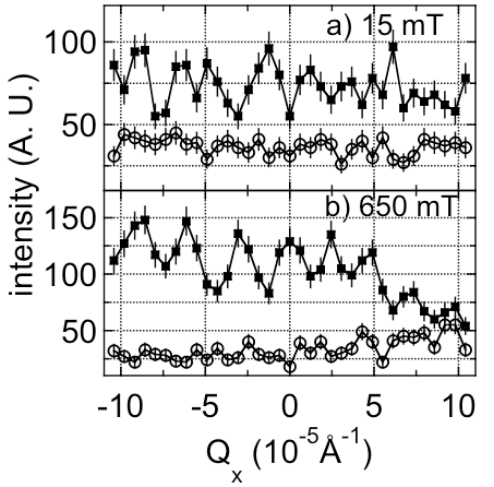}
\caption{Field-dependent polarized neutron transverse scans at \textit{Q}$%
_{z}$ = 0.035$%
\operatorname{\text{\AA}}%
^{-1}$. Error bars correspond to $\pm$$\sigma$. Lines are guides to eye.
\label{PNR}}
\end{figure}

Having obtained the effect of Pr-doping on the interfacial electronic
structure, we also performed neutron reflectometry measurements, which are sensitive to the
depth-dependent and planar composition (structural and magnetic) of
thin-film multilayers.\cite{Majkrzak,Zabel} This was for
comparison to previous NR studies on YBCO/LCMO multilayers grown on STO which
exhibited interesting oscillatory diffuse scattering as SC
sets in\cite{Chakhalian1} below the STO cubic-to-tetragonal phase transition
at T$_{STO}$ = 105 K.\cite{Vlasko-Vlasov}
NR measurements were performed at the NIST Center for Neutron Research,
using the AND/R and NG-1 reflectometers.\cite{Dura,web} For all
measurements, the sample was mounted using a flexible aluminum backing (to minimize stress), and was cooled from room temperature to 150 K in zero
field, and then further cooled to 7 K in the presence of an applied magnetic
field. Figure 3 shows unpolarized neutron scattering reciprocal space maps
measured by a position sensitive detector. Scattering along the z-component of wavevector
transfer (\textit{Q}$_{z}$) (specular scattering) originates from structural
and magnetic features along the growth direction of the sample, while
scattering along the \textit{Q}$_{x}$ axis (diffuse scattering) corresponds
to in-plane features. At room temperature (3a), purely nuclear scattering is
observed along the specular ridge (\textit{Q}$_{x}$=0) and a superlattice Bragg
peak is clearly
observable (\textit{Q}$_{z}$ $\approx$ 0.035$%
\operatorname{\text{\AA}}%
^{-1}$). As the temperature is reduced below T$_{STO}$ (3b-c), the
specular ridge dramatically diffuses out along the \textit{Q}$_{x}$ direction- indicating
significantly increased in-plane inhomogeneity. To identify the nature of
this planar inhomogeneity, a series of higher resolution \textquotedblleft
transverse scans\textquotedblright\ (fixed \textit{Q}$_{z}$) were taken
using a 3He pencil detector. Figure 4 shows temperature-dependent,
unpolarized beam transverse scans taken at \textit{Q}$_{z}$ = 0.0139$%
\operatorname{\text{\AA}}%
^{-1}$ (just below the critical edge) in a 650 mT field. Below T$_{STO}$,
diffuse scattering is again observed, but in this case pronounced
oscillations in \textit{Q}$_{x}$ are clearly resolvable.\cite{oscillations}
Qualitatively similar results were obtained at a 15 mT field (not shown). Figures 4b-d show that this oscillatory
diffuse scattering is present both above and below T$_{SC}$, indicating that the onset of
SC plays no role in its origin. Figure 5 shows the results of
polarized beam transverse scans taken at the 1st superlattice Bragg position
after cooling in 650 mT (5a) and 15 mT (5b). Non spin-flip
scattering of spin-up or spin-down neutrons is evident (no significant
spin-flip scattering could be detected). While the overall spin-splitting
(related to the total sample magnetization\cite{Majkrzak,Zabel}) increases
with increasing field, the oscillatory
nature of the diffuse scattering is similar for both fields - strongly
suggesting that it is not a magnetic effect. Instead, it is exceedingly
likely that the observed in-plane modulation described above is primarily
due to the STO structural phase transition which is associated with
crystallographic twinning and surface buckling.\cite{Chakhalian1,Vlasko-Vlasov} Specifically, we expect that the superlattice
film becomes faceted below T$_{STO}$, causing the specular reflection to
split into multiple reflections, similar to the observation by Hoppler et
al. (Ref. \onlinecite{Hoppler1,Hoppler2}). A SC-induced
period doubling of the magnetization profile along the c-axis was not
observed in our study, perhaps as a consequence of the previously noted\cite%
{Hoppler1} sensitivity of this phenomenon to external boundary conditions.

In conclusion, we have grown high-quality Y$_{0.6}$Pr$_{0.4}$Ba$_{2}$Cu$_{3}$O$_{7}$/La$%
_{2/3}$Ca$_{1/3}$MnO$_{3}$ superlattices.The interfacial reconstruction
was studied by linearly polarized X-ray absorption at the Cu L$_{3}$-edge.
Similar to the interface between optimally doped YBCO and LCMO, we observed
dramatic modifications including charge transfer and OR
at the interfacial CuO$_{2}$ plane of Pr-doped YBCO due to the strongly
hybridized covalent bond between Cu and Mn. However, differences such as
reduced charge transfer and more bulk-like orbital occupation were also
found. This is presumably related to the Pr-doping which removes holes from
the ZR state and suppresses SC. Neutron reflectivity
data revealed temperature-dependent, oscillatory diffuse scattering,
indicative of a significant in-plane modulation, which can be attributed to
buckling of the superlattice film due to the structural phase transition of
the STO substrate.

\textit{Acknowledgments} J.C. was supported by DOD-ARO under the Contract
No. 0402-17291 and NSF Contract No. DMR-0747808. Work at the Advanced Photon
Source, Argonne is supported by the U.S. Department of Energy, Office of
Science under Contract No. DEAC02-06CH11357.

\end{document}